\documentclass[aps,prl,
 reprint,
 superscriptaddress,
 showpacs,
 amsmath,amssymb,
 aps,
]{revtex4-1}

\usepackage{graphicx}
\usepackage{dcolumn}
\usepackage{bm}
\usepackage{hyperref}

\usepackage{epstopdf}
\usepackage{todonotes}
\usepackage{siunitx}
\usepackage{units}
\newcommand{\p}{\hphantom{\ensuremath{-}}}
\newcommand{\pd}{{\phantom{\dagger}}}

\begin{document}


\title{Realistic quantum critical point in one-dimensional two-impurity models}

\author{Benedikt Lechtenberg}
\email{benedikt.lechtenberg@tu-dortmund.de}
\author{Fabian Eickhoff}
\author{Frithjof B. Anders}
\affiliation{Lehrstuhl f\"ur Theoretische Physik II, Technische Universit\"at Dortmund, 44221 Dortmund,Germany}

\date{\today}

\begin{abstract}
We show that the two-impurity Anderson model exhibits an additional quantum critical point 
at infinitely many specific distances between both impurities for an inversion symmetric one-dimensional dispersion.
Unlike the quantum critical point previously established,
it is robust against particle-hole or parity symmetry breaking.
The quantum critical point
separates a spin doublet from a spin singlet ground state and is, therefore, protected.
A finite single-particle tunneling $t$ 
or an applied uniform gate voltage will drive the system across the quantum critical point.
The discriminative magnetic properties of the different phases cause a jump in the spectral functions at low
temperature, which might be useful for future spintronics devices.
A local parity conservation will prevent the spin-spin correlation function from decaying to its
equilibrium value after spin manipulations.

\end{abstract}


\maketitle

\paragraph{Introduction.}
The promising perspective of combining traditional electronics with novel spintronics devices leads to intense research into controlling and switching the magnetic properties of such nanodevices. 
Experimentally magnetic properties of adatoms on surfaces 
\cite{lit:RevModPhys.76.323,lit:Spintronic_anisotropy_Misiorny2013,lit:Graphene_spintronics_Han2014,lit:Metals_on_silicon_Johll2014,lit:Defect_graphene_Yazyev2007,lit:Bork_Kroha2011}
or magnetic molecules 
\cite{lit:lechtenberg_2016_dimer,lit:spinfilter2010,lit:spintronic_review_Bogani2008,lit:Chemistry_Review_spintronics_Sanvito2011,lit:Review_Naber2007,lit:Organic_Dediu2009, lit:spin_diffusion_Drew2009,lit:Spinelli2015}
might serve as the smallest building
blocks for spintronic devices.

From a theoretical perspective,
the two-impurity Anderson model (TIAM) \cite{Doniach77,lit:Jabben2012,lit:Sakai1992_I}
constitutes an important but simple system  which embodies
the competition of interactions between two 
localized magnetic moments with those between the impurities and the conduction band.
The TIAM has been viewed as a paradigm model for the formation
of two different singlet phases separated by a quantum critical point (QCP):
a Ruderman-Kittel-Kasuya-Yosida (RKKY)-induced singlet  and a Kondo singlet \cite{Doniach77}.
This quantum critical point (QCP) 
investigated by Jones and Varma \cite{lit:Jones88,lit:Jones1989,lit:Sakai1992_I}, however,
turned out to be unstable against particle-hole (PH) symmetry breaking \cite{lit:Affleck95} 
and the two different singlet phases are adiabatically connected. 
This led to the conclusion that for finite distances between the impurities, no QCP exists and the original finding is just a consequence of unphysical approximations \cite{lit:Fye1994}
which is generically replaced by a crossover regime. 

In this paper, we establish that the model exhibits another realistic QCP for any inversion symmetric one-dimensional (1D) dispersion, depending only on the absolute value of the wave vector.
The existence of this different QCP relies only on the fact that for specific distances $R$ between 
both impurities, either the even- or odd-parity contributions to the conduction band
decouple from the impurities at low-energy scales, leading to an underscreened Kondo effect.
This underscreened Kondo fixed point (USK FP) \footnote{
The USK FP differs from the strong-coupling fixed point
usually discussed in the context of an underscreened Kondo effect by an additional
free conduction band with zero phase shift.
}
has a doublet ground state which
is different from
the singlet ground state for large antiferromagnetic interactions between both impurities, 
excluding a smooth crossover between both phases.
This QCP trivially also exists for the limit $R\to 0$ in all dimensions \cite{lit:R_zero_QPT_Vojta_Bulla_Hofstetter2002,lit:NishimotoPruschke2006}: 
For this special case, the QCP has been recently observed 
in molecular dimers \cite{lit:lechtenberg_2016_dimer}, where
the different phases can clearly be detected in the scanning tunneling spectra.

Here, we present the generalization to finite distances,
its robustness against particle-hole symmetry as well as parity breaking, and 
demonstrate that the quantum phase transition (QPT) 
can also be evoked by applying a gate voltage to the impurities.
Since the entanglement between the impurity spins is protected by a dynamical
symmetry in the parity-symmetric case, the spin-spin correlation function cannot completely 
decay to its equilibrium value, and, therefore, might be useful for future qubit implementations.
Possible experimental realizations for finite distances could be in pseudo-1D nanostructures
\cite{lit:Kondo_one_dimension_Blachly_1992,lit:Finite-size_Kondo_DiTusa_1992,lit:Kondo_reduced_dimensionality_Blachly_1995,lit:Kondo_wires_Mohanty_2000,lit:Synthesis_one_dimensional_metal_with_insulated_molecules_2014}
or optical lattices \cite{lit:Gorshkov2010,lit:Controlling_ultracold_atoms_Kondo_Duan_2004,lit:Fermionic_atoms_optical_lattice_Paredes_2005}.

\paragraph{Model.}
We consider the two-impurity Anderson model (TIAM) whose Hamiltonian can be separated into the parts $H_\mathrm{TIAM}=H_\mathrm{c} + H_\mathrm{D} + H_\mathrm{I}$.
$H_\mathrm{c}$ contains the conduction band $H_\mathrm{c}=\sum_{\vec{k},\sigma}\epsilon({\vec{k}})c^{\dagger}_{\vec{k},\sigma}c^{\p}_{\vec{k},\sigma}$
and $H_\mathrm{D}$ and $H_\mathrm{I}$ comprise the impurity contribution and the interaction between the conduction band and impurities, respectively,
\begin{align}
	H_\mathrm{D} =&\sum_{j,\sigma} E_j d^\dagger_{j,\sigma} d^{\phantom{\dagger}}_{j,\sigma} + U \sum_j n_{j,\uparrow}n_{j,\downarrow} + \vec{h}\sum_j \vec{S}_j \nonumber \\
	    &+ \frac{t}{2} \sum_\sigma \left( d^\dagger_{1,\sigma} d^{\phantom{\dagger}}_{2,\sigma} + d^\dagger_{2,\sigma} d^{\phantom{\dagger}}_{1,\sigma} \right) \label{eq:TIAM_D_12} \\
	H_\mathrm{I}= &\frac{V}{\sqrt{N}}\sum_{j\in\{1,2\}k,\sigma} c^\dagger_{{k},\sigma} e^{\mathrm{i}{k}{R}_j} d_{j,\sigma} + \mathrm{h.c.} , \label{eq:TIAM_I_12}
\end{align}
with $d^\dagger_{j,\sigma}$ creating an electron with spin $\sigma$ and energy $E_j$ 
on impurity $j$ located at position $R_{1/2}=\pm R/2$, $n_{j,\sigma}=d^\dagger_{j,\sigma} d^{\p}_{j,\sigma}$, 
a local magnetic field $\vec{h}$ applied to the spin $\vec{S}_j=\frac{1}{2} d^\dagger_{j,\sigma} \vec{\sigma}_{\sigma,\sigma'}d^{\phantom{\dagger}}_{j,\sigma}$ of impurity $j$,
and $c^\dagger_{\vec{k},\sigma}$ creating a conduction electron.
At low temperatures, the tunneling $t$ leads to an effective antiferromagnetic exchange interaction $K\vec{S}_1\vec{S}_2$, with $K=t^2/U$ between the impurity spins.
Throughout this work, unless stated otherwise, we will consider the case $E_1=E_2=E=-U/2$
for simplicity such that  both impurities are occupied with one electron.
Below, we will show that the QCP is wholly robust to a departure from parity and particle-hole symmetries.

For the numerical renormalization-group (NRG) approach \cite{lit:WilsonNRG,lit:Krishnamurthy08_I,lit:Krishnamurthy08_II,lit:BullaReview}, 
it is useful 
to introduce a parity eigenbasis $d_{e/o,\sigma} =\frac{1}{\sqrt{2}}( d_{1,\sigma} \pm d_{2,\sigma} )$ for the impurity degrees of freedom
\cite{lit:JayaprakashKrischnamurtyWilkins1981,lit:Jones87,lit:Jones88,lit:Jones1989,lit:Affleck95, lit:Borda07,lit:lechtenbergAnders14,lit:lechtenberg_2016_dimer}.
In this basis, the orbitals with even/odd parity couple to corresponding even/odd parity conduction bands via the energy- and distance-dependent hybridization functions (see Supplemental Material),
\begin{subequations}
\begin{align}
\Gamma_{e}(\epsilon,{\vec{R}})= &  \frac{2\pi V^2}{N} \sum_{\vec{k}} \delta(\epsilon-\epsilon(\vec{k})) \cos^2\left(\frac{\vec{k}\vec{R}}{2}\right) \\
\Gamma_{o}(\epsilon,{\vec{R}})=&  \frac{2\pi V^2}{N} \sum_{\vec{k}} \delta(\epsilon-\epsilon(\vec{k})) \sin^2\left(\frac{\vec{k}\vec{R}}{2}\right). 
\end{align}
\label{eq:HybridizationFunctions}
\end{subequations}
A proper consideration of the energy dependence of these functions generally breaks 
particle-hole symmetry \cite{lit:Fye1994,lit:Affleck95} and hence destroys the well-known QCP predicted by Jones and Varma \cite{lit:Jones87,lit:Jones88,lit:Jones1989}.

\paragraph{Hybridization functions.}
Examining the definitions of the hybridization functions $\Gamma_{e/o}(\epsilon,{\vec{R}})$
reveals an important fundamental property:
If all wave vectors $\vec{k}'$ fulfilling $\epsilon({\vec{k}'})=0$ also satisfy the condition 
${\vec{k}'}{\vec{R}_n}=\pi n$, 
with $n$ being an integer, 
one of the two hybridization functions  exhibits a pseudogap 
$\propto |\epsilon|^2$
because either the sine or the cosine in Eqs. \eqref{eq:HybridizationFunctions} vanishes for $\epsilon \to 0$.
While for a general dispersion this requirement is not fulfilled, 
infinitely many equidistant   $R_n=|\vec{R}_n|$ 
obeying this requirement are found for
a 1D inversion symmetric dispersion 
with $\epsilon(k)=\epsilon(|{k}|)$.
Note that the presented results are valid for the case that the mean free path of the electrons in the conduction channel is larger than the distance $R$.

Since the Kondo screening breaks down for a pseudogap hybridization function vanishing as $|\epsilon|^r$,
with $r>1/2$ \cite{lit:Bulla97,lit:Ingersent1998,lit:ChenJayaprakash1995,lit:Ingersent1996},
the Kondo effect of the even or odd conduction band will disappear for the specific distances ${k'}{R_n}=\pi n$,
leading to an underscreened spin-$1$ Kondo fixed point (USK FP)
with an effective free spin-$1/2$ remaining.

The odd-hybridization function completely vanishes for any dispersion and $R\to 0$ on all energy scales, leading to a single-channel model and, thus, trivially to an USK FP.
For a 1D linear dispersion $\epsilon(k)=v_\mathrm{F}(|k|-k_\mathrm{F})$,
Eqs.\ \eqref{eq:HybridizationFunctions} 
yield \cite{lit:Borda07,lit:lechtenbergAnders14}
\begin{align}
	\Gamma_{e/o}^{\mathrm{1D}}(\epsilon,{R})=\Gamma_0\left( 1 \pm \cos\left[ k_\mathrm{F} R ( 1 + \frac{\epsilon}{D})\right] \right) \label{eq:HybridizationFunctions_1D}
\end{align}
with $\Gamma_0=\pi\rho_0 V^2$, the half bandwidth $D$, the constant density of states of the original conduction band $\rho_0=1/2D$, $k_\mathrm{F}=\pi/2a$, and $a$ the lattice constant.
The hybridization function of the even conduction band exhibits a gap for distances $k_\mathrm{F}R=(2n+1)\pi$ and the one of the odd band for $k_\mathrm{F}R=2n\pi$.
Note that with increasing distance $R$, the frequency of the oscillations in $\Gamma^\mathrm{1D}_{e/o}(\epsilon,{R})$ increases and, consequently,
the width of the gap becomes smaller so that the 
stable low-energy FP is reached at increasingly lower temperatures.

\begin{figure}[t]
	\includegraphics[width=0.5\textwidth]{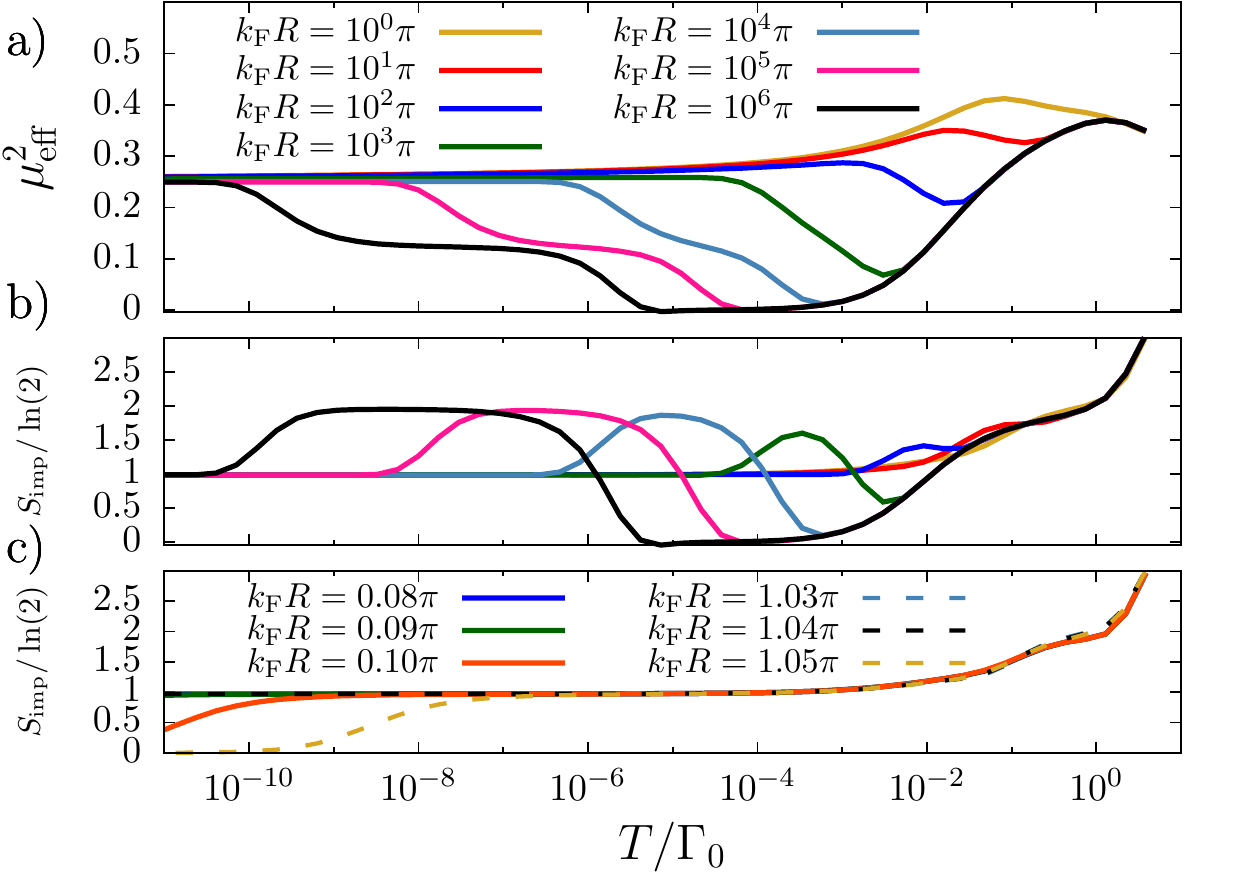}
	\caption{(a) The effective local magnetic moment $\mu^2_\mathrm{eff}$ and (b) entropy of the impurities for the TIAM plotted against the temperature $T$ for different distances.
		(c) The temperature-dependent entropy for distances that slightly deviate from $R_n=0$ (solid lines) and $R_n=1$ (dashed lines).
		Model parameters are $t=0$, $E=-5\Gamma_0$, $U=10\Gamma_0$, and $D=10\Gamma_0$.
	}
	\label{fig:J06_mu_entropy}
\end{figure}

\paragraph{Doublet ground state.}
Generically, a singlet ground state is found in the TIAM
since either the two impurity spins are bound in a local singlet for strong antiferromagnetic 
correlations between the impurities 
or the impurity spins are screened by 
the surrounding conduction band electrons to spatially extended Kondo singlets \cite{lit:Jones87,lit:Jones88,lit:Jones1989}.
A different situation arises for the specific distances $k_\mathrm{F}R_n=n\pi$,
where one conduction band decouples at low energies.
This is demonstrated in Fig.\ \ref{fig:J06_mu_entropy} where 
the effective impurity magnetic moment $\mu^2_\mathrm{eff}$ 
and the entropy $S_{\rm imp}$  \footnote{Both quantities $\mu^2_\mathrm{eff}$ and $S_{\rm imp}$ are calculated as usual in the NRG via $A(T)=A_\mathrm{tot}(T) - A_\mathrm{free}(T)$, 
where $A_\mathrm{tot}(T)$ is the measured quantity of the total system, consisting of the impurities coupled to the bath.
From this, the quantity of a reference system, i.e., one without impurities, is subtracted.}
are plotted for different $R_n$.
The USK FP with a free unquenched spin-$1/2$ remaining
is the only stable fixed point for vanishing spin-spin interaction $K=0$ ($t=0$)
characterized by  $\mu^2_\mathrm{eff}=0.25$ and the impurity entropy $S_\mathrm{imp}=\ln(2)$.

At very large distances $R_n$, the gap in one of the hybridization functions becomes very narrow so that the 
crossover to the USK FP only occurs at very low temperatures.
For such distances, at first both impurities are screened by the two conduction bands, leading to an almost vanishing magnetic moment $\mu^2_\mathrm{eff}\approx 0$ and entropy $S\approx 0$.
However, the renormalization of the effective Kondo coupling
and consequently the screening of one local spin always stops at a finite temperature 
due to the pseudogap hybridization function
and, therefore, the screening is never complete.
Since the hybridization to one conduction band vanishes at the Fermi energy, the coupling to 
that band subsequently decreases
until finally the USK FP emerges at very low temperatures.

In between these two FPs,
the model exhibits another unstable FP
with $\mu^2_\mathrm{eff}=0.125$ and entropy $S_\mathrm{Imp}=2\ln(2)$.
The values for $\mu^2_\mathrm{eff}$ and $S$ are a feature of the
the gapped Wilson chain  \footnote{More details on the origin of this unstable fixed point can be found in the Supplemental Material.}
and are not related to the impurity physics.
While $\mu^2_\mathrm{eff}(T)$ starts to increase until it reaches the value $\mu^2_\mathrm{eff}=0.125$ in the regime of the unstable FP,
the impurity spins remain screened so that the local moment of the impurities $\mu^2_\mathrm{loc}(T) = T \lim_{h_z\to 0} \langle S^z_j\rangle/h_z$ \cite{lit:Ingersent1998,Chowdhury2015}
continues to decrease linearly with decreasing $T$.
Since the impurity spins are only completely screened at  $T=0$ in the conventional Kondo problem, 
the screening of the impurity spins progresses until the USK FP is reached at low temperatures where the local moment $\mu^2_\mathrm{loc}(T)$ 
and remains constant for $T\to 0$
as it is expected for a free but strongly reduced magnetic moment 
in the Curie-Weiss law \footnote{
The temperature dependent behavior of $\mu^2_\mathrm{loc}(T)$ is depicted in the Supplemental Material.
}.

\begin{figure}[t]
	\includegraphics[width=0.4\textwidth,height=130px]{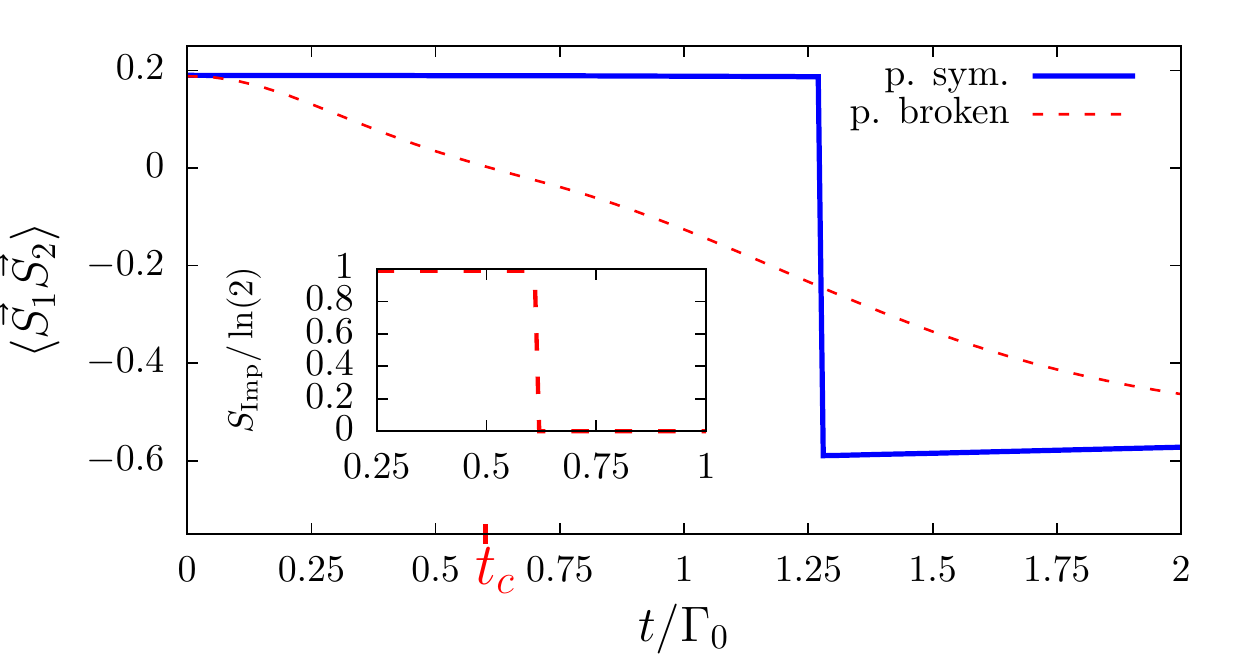}
	\caption{The correlation function $\langle\vec{S}_1\vec{S}_2\rangle$ plotted against the tunneling $t$ for the distance $k_\mathrm{F}R=\pi$.
	 While in the parity-symmetric cases (solid lines) $\langle\vec{S}_1\vec{S}_2\rangle$ must change discontinuously,
	in the parity-broken case the correlation function is continuous. 
	The inset shows the entropy for the parity-broken case and the new critical $t_c(\Delta E)$ is marked on the $x$-axis.
	}

	\label{fig:S1S2}
\end{figure}

The low-temperature crossover scale from the unstable FP to the stable USK FP
depends on the degree of screening: the smaller $\mu^2_\mathrm{loc}(T_\mathrm{Gap})$ at the energy scale $T_\mathrm{Gap}$ at which the pseudogap develops,
the smaller the crossover temperature scale.
Such a vigorous screening can be achieved in two ways: either the distance $R_n$
is increased so that the screening stops at lower temperatures (shown in Fig. \ref{fig:J06_mu_entropy})
or the coupling  $V$ to the bands is increased so that the impurities are already strongly screened at a higher $T$.

Although a small departure from the specific distances $R_n$ theoretically always leads to a singlet ground state at very low temperatures,
the system will stay in the now unstable doublet fixed point for all experimentally relevant temperatures if the departure is not too large, which can be seen in Fig. \ref{fig:J06_mu_entropy}(c).
As a result, using $k_\mathrm{F}=\pi/(2a)$ of a linear dispersion yields that in experiments at finite temperatures, variances of $R_n$ by up to $20\;$\% of the lattice constant are still sufficient
to detect a sharp change in the magnetic properties of the system.

\paragraph{Quantum critical point.}
While for a vanishing spin-spin interaction between the impurities the ground state is always a doublet at $R_n$, 
both impurity spins form a local spin singlet for sufficiently strong antiferromagnetic interactions $K$.
Therefore, these two phases must be separated by a QCP.
Unlike the unstable Jones-Varma QCP \cite{lit:Jones88,lit:Jones1989,lit:Affleck95} separating 
two singlet ground states, this QCP is protected 
by the spin as a conserved quantum number.
While the Jones-Varma QPT is continuous \cite{lit:Jones1989,lit:Affleck95},
we found in the parity-symmetric case a linear decreasing energy scale with decreasing $|t-t_c|$ typical for a parity-protected level-crossing QPT,
while in the parity-broken case we observed a exponentially vanishing energy scale indicating 
a Kosterlitz-Thouless QPT \cite{lit:Mitchell_2013,lit:Mitchel_2009,lit:R_zero_QPT_Vojta_Bulla_Hofstetter2002,lit:Hofstetter2001}.

\begin{figure}[t]
	\includegraphics[width=0.5\textwidth]{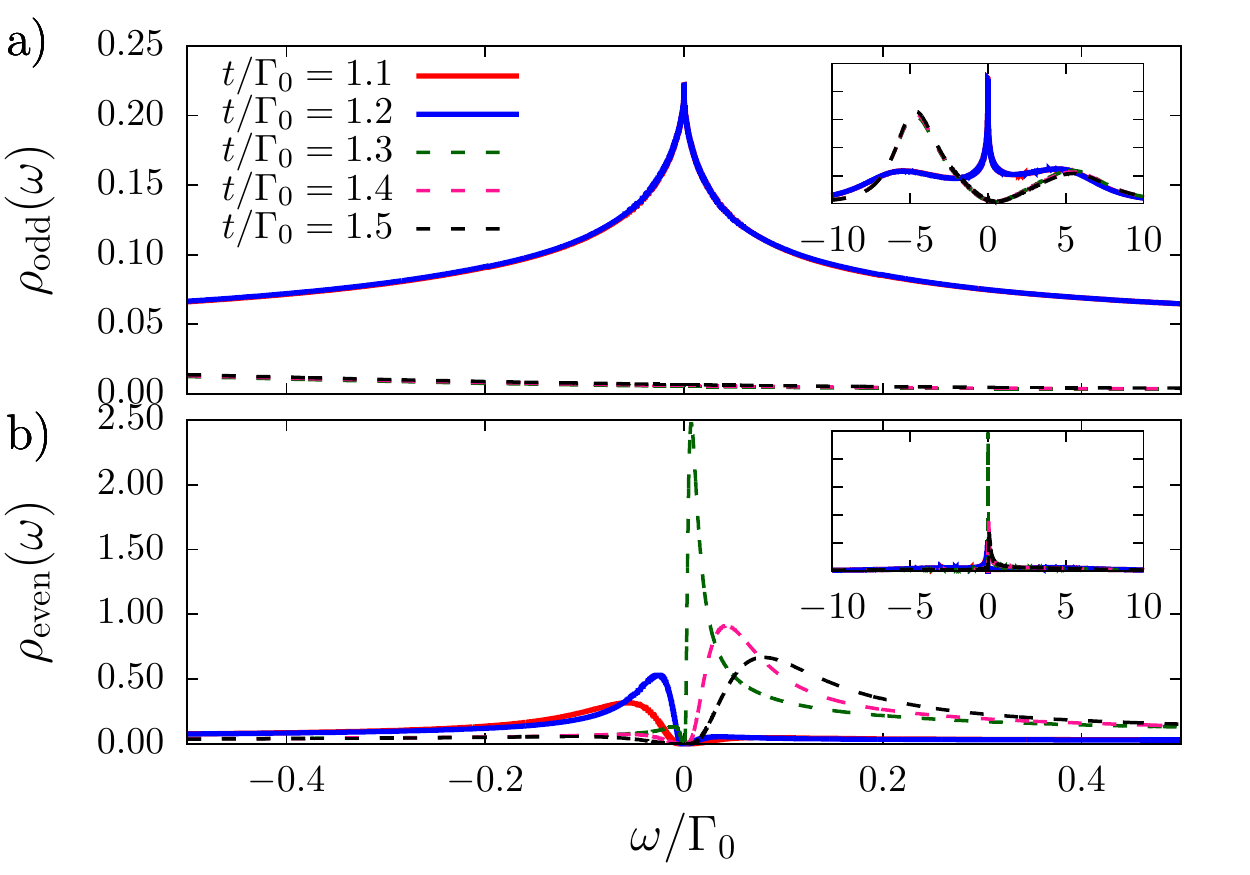}
	\caption{Spectral function of the (a) odd and (b) even orbital for tunnelings $t<t_c$ (solid lines) and $t>t_c$ (dashed lines)
	and $\Delta E=0$.
	  For the distance $k_\mathrm{F}R=\pi$, the even orbital decouples for $T\to 0$ from the conduction band.
	}
	\label{fig:SpectralFunctions}
\end{figure}

The different nature of the QPTs is also revealed in the local correlation function $\langle \vec{S}_1\vec{S}_2 \rangle$.
While in the absence of $K$ ($t=0$) a local triplet screened by the Kondo effect at low $T$
to a doublet is part of the ground state, 
a local singlet forms and 
suppresses the Kondo effect \cite{lit:Jones88,lit:Jones1989} for large antiferromagnetic $K$.
This leads to  $\langle \vec{S}_1\vec{S}_2 \rangle > 0$ in the former regime, while in the latter, one finds $\langle \vec{S}_1\vec{S}_2 \rangle < 0$.
For the Jones-Varma QCP, $\langle \vec{S}_1\vec{S}_2 \rangle$
varies continuously across the QPT and only its derivative diverges at the QCP.
This is due to a mixing term in the Hamiltonian \cite{lit:Jones88,lit:Jones1989,lit:Sakai1992_I} exchanging
even and odd conduction electrons via impurity scattering processes. 
Global parity remains conserved, but local parity on the impurity subsystem is broken. 
This is contrasted by the behavior of $\langle \vec{S}_1\vec{S}_2 \rangle$ at $R_n$, as shown in Fig.\ \ref{fig:S1S2}. 
Since one band decouples at low-energy scales, the band mixing term is suppressed
and a dynamical local parity conservation 
ensures the conservation of $\langle \vec{S}_1\vec{S}_2 \rangle$ at low temperatures
and prohibits the decay to its equilibrium value after a spin manipulation.
Consequently, the correlation function has to change discontinuously at the QCP for a parity-symmetric model \footnote{
For a more detailed explanation of the origin of the discontinuity, we refer the reader to the Supplemental Material,
which includes Refs. \cite{lit:Galpin2005,lit:SchriefferWolff1966, lit:ZitkoBonca2006}.
}.

Furthermore, the QPT is even robust against parity breaking: We have added a small $\Delta E$ to one of
the two single-particle levels, i.e.,  $E_1=E+\Delta E$, which is one of several ways of breaking the parity.
Although the spin-correlation function varies now continuously in the parity-broken case, as depicted in Fig.\ \ref{fig:S1S2},
other quantities such as the magnetic moment $\mu^2_\mathrm{eff}$, 
the entropy $S_\mathrm{imp}$ (shown in the inset of Fig.\ \ref{fig:S1S2}), or the spectral functions still show a discontinuity 
at the renormalized critical tunneling $t_c(\Delta E)$, marked on the $x$-axis in Fig.\ \ref{fig:S1S2}.

For the parity-conserving case, the spectra of the odd and even orbital
\cite{lit:Pruschke06,lit:Weichselbaum_Delft2007}
are shown in Fig.\ \ref{fig:SpectralFunctions} for the two different phases and the distance $k_\mathrm{F}R=\pi$, 
at which the even orbital decouples from the conduction band at low-energy scales.
The spectral functions exhibit the same features as in the $R=0$ case \cite{lit:lechtenberg_2016_dimer,lit:NishimotoPruschke2006},
but with the role of even and odd spectra interchanged.

The spectrum for the odd orbital develops an underscreened Kondo peak \cite{lit:Florens_Costi2009} at $\omega=0$ 
for $t<t_c$,
which collapses once the tunneling exceeds  $t>t_c$. 
In this phase, both impurity spins are bound into a local singlet.

In contrast, $\rho_{\rm even}$ always develops a gap around the Fermi energy for all $t\not = t_c $:
the  pseudogap in the even-hybridization function suppresses the Kondo screening of the spin in the even orbital. 
Furthermore, at low frequencies, the orbital decouples from the hybridization processes.
Injecting/ejecting an electron into/from the even orbital changes the local particle number, which cannot relax but
induces a suddenly changed Coulomb potential for the odd orbital. 
The only way the system can respond at $T = 0$ is by changing the many-body ground state.
This leads to the well-understood x-ray edge physics \cite{RouletGavNozieres69,*NozieresDeDominicis1969} also found in the Falicov-Kimball model \cite{lit:Kotliar1992,*AndersCzycholl2005}.
The excitations around the Fermi energy thus indicate transitions from the doublet to the singlet phase for $t<t_c$, and vice versa for $t>t_c$.
Consequently, the width of the gap in the spectrum is given by the energy difference between the doublet and singlet state and vanishes for $t\to t_c$.
Note that for distances $k_\mathrm{F}R=2n\pi$, the spectral functions of the even and odd are interchanged.

In the general parity-broken case, features of the even orbital are weakly mixed into the spectral function of the odd orbital, and vice versa, 
since in this case both orbitals are coupled to both conduction bands \cite{lit:lechtenberg_2016_dimer}.
Experimentally, the QPT can be detected by measuring the differential conductance though an impurity which is proportional to a superposition of the even and odd spectral functions.
We predict that  for $t<t_c$, a clear Kondo peak at the Fermi energy is visible 
below the Kondo temperature $T_\mathrm{K}$.
This Kondo peak disappears for $t>t_c$, and only the finite frequency excitations
stemming from the x-ray edge physics of the weakly coupled orbital
are mixed in as recently detected in a molecular dimer system\cite{lit:lechtenberg_2016_dimer}.

\begin{figure}
	\centering
	\includegraphics[width=0.45\textwidth]{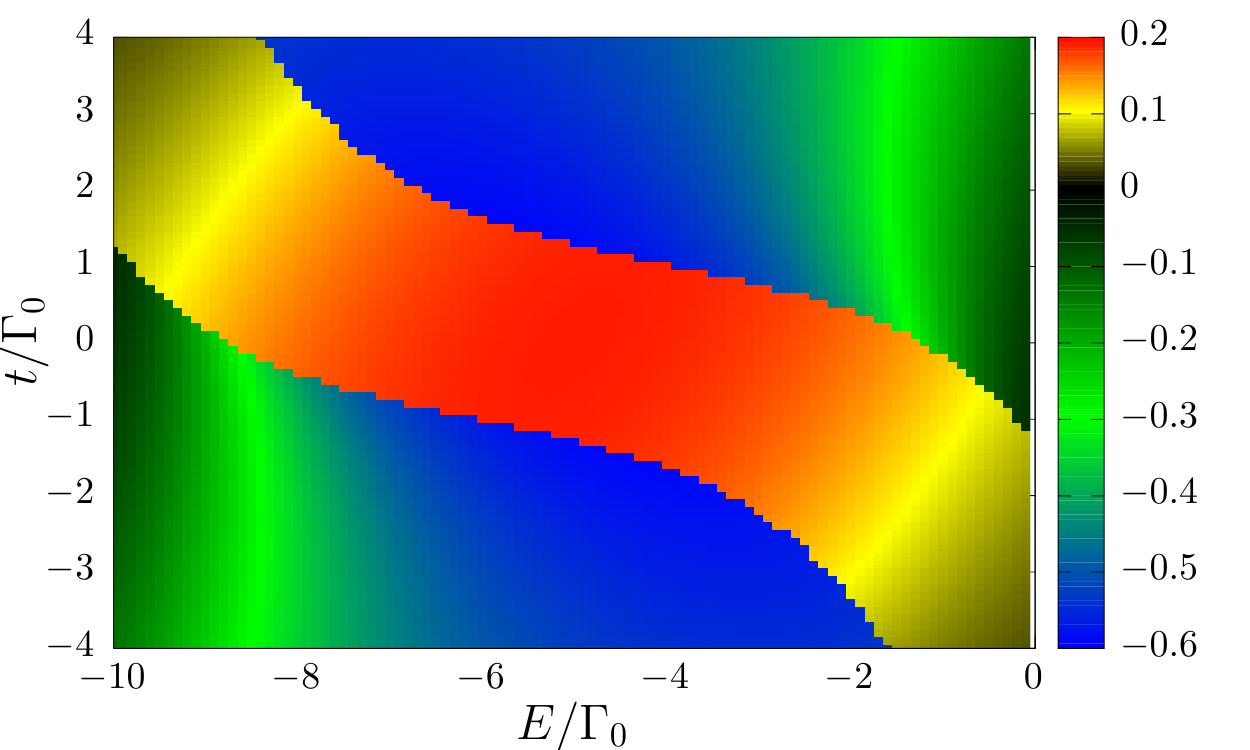}
	\caption{$\langle \vec{S}_1\vec{S}_2\rangle$ phase diagram plotted against $E$ and $t$ for $k_\mathrm{F}R=\pi$ and $U=10\Gamma_0$. 
		}
	\label{fig:phasediagram}
\end{figure}

Since the tunneling $t$ is generated by the overlap of orbital wave functions of the adatoms or molecules in experiment \cite{lit:lechtenberg_2016_dimer}, variation of the tunneling $t$ is experimentally difficult. 
The case of a fixed $E$ but different discrete $t$ changed via molecule geometry 
has been recently realized \cite{lit:lechtenberg_2016_dimer} for the extreme case of $R\approx 0$, but is not suitable for
electronic switching of the local spin configuration.

However, it is also possible to evoke the QPT for a fixed tunneling $t$ via a gate voltage shifting both orbital level energies $E$.
Figure \ref{fig:phasediagram} depicts a phase diagram of the correlation function $\langle \vec{S}_1\vec{S}_2\rangle$ as a function of $E$ and $t$.
Ferromagnetic correlations (red and yellow), indicating that the system is in the doublet phase, 
are developing inside a tube.
If either the tunneling $t$ or the energy level $E$ is sufficiently increased or decreased, 
the system is found in the singlet phase (blue and green).
Inside the tube, the local magnetic moment and the impurity entropy take the fixed values 
$\mu_\mathrm{eff}^2=0.25$ and $S_\mathrm{imp}=\ln(2)$, while outside both vanish \footnote{
The phase diagrams of $\mu_\mathrm{eff}^2$ and $S_\mathrm{imp}$ are shown in the Supplemental Material.
}.
For very large positive or negative level energies, $|\langle \vec{S}_1\vec{S}_2\rangle|\to 0$ decreases continuously since the orbitals become either doubly occupied or empty.
Note that in this case, the Kondo effect will also break down in the doublet phase since there is no local moment in the coupled orbital that can be screened.
To understand the asymmetry with respect to $E$ and $t$, it is useful to monitor the single-particle energies in the even-/odd-parity basis where both energies are split by the tunneling $E_{e/o}=E\pm t/2$
so that the even/odd level energy is increased/decreased with increasing $t$.
In order to evoke a transition from the singlet to the doublet phase, the decoupled orbital has to be shifted towards half filling such that it becomes singly occupied again, which can only happen discontinuously.
Consequently, if the distance is changed from an odd distance $k_\mathrm{F}R/\pi=2n+1$, shown in Fig. \ref{fig:phasediagram}, 
to an even distance, the roles of the even/odd orbital as the uncoupled/coupled orbital are interchanged and the phase diagram is hence mirrored at the line $t=0$.

\paragraph{Summary.}
We have shown that the TIAM exhibit a QCP for a 1D dispersion $\epsilon(k)=\epsilon({|k|})$ in the cases that the impurities are separated by specific distances $R_n$.
In contrast to the unstable QCP \cite{lit:Jones88,lit:Jones1989} 
usually discussed in the context of the two-impurity models,
the QCP presented in this paper
is stable to departure from particle-hole and parity symmetry.

We believe that this system may be of great relevance for spintronic devices since it is possible by applying gate voltages to turn on and off a free magnetic moment which is not screened at low temperatures.
Along with the magnetic moment, one can switch on and off a Kondo effect with its sharp conductance peak at the Fermi energy.
Furthermore, in the parity-symmetric case, the spin-spin correlation between both impurity spins is protected by the parity as a conserved quantity,
making this system promising for spin-qubit realizations.

\begin{acknowledgments}
We acknowledge useful discussions with S.\ F.\ Tautz and R.\ Bulla.
B.L. thanks the Japan Society for the Promotion of
Science (JSPS) and the Alexander von Humboldt Foundation.
\end{acknowledgments}

%

\widetext
 \newpage
\begin{center}
 \Large{\textbf{Supplemental Material}}
\end{center}


\renewcommand{\theequation}{S\arabic{equation}}
\renewcommand{\thefigure}{S\arabic{figure}}

\setcounter{equation}{0}
\setcounter{figure}{0}
%

\section{Mapping to the parity basis}
\begin{figure}[t]
	\includegraphics[width=0.65\textwidth]{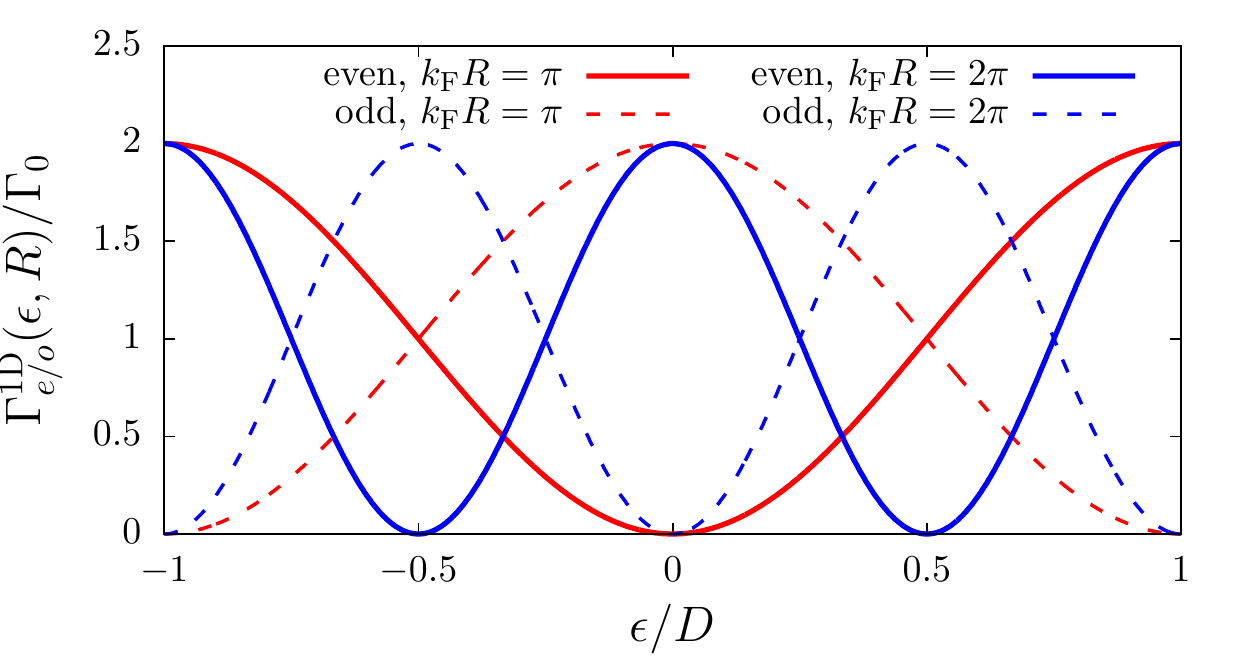}
	\caption{
		Hybridization functions of Eq. \eqref{Seq:HybridizationFunctions_1D} for a linear dispersion in 1D for two different distances $k_\mathrm{F}R=\pi$ (red) and $k_\mathrm{F}R=2\pi$ (blue).
		For these differences either the even (solid) or the odd (dashed) hybridization function exhibit a pseudogap leading to a breakdown of the Kondo effect of the associated conduction band.
	}
	\label{fig:DOS}
\end{figure}
While Wilsons original numerical renormalization group (NRG) approach was only designed to solve the thermodynamics of one localized impurity, 
the NRG was later successfully extended by Jones and Varma \cite{Slit:Jones87,Slit:Jones88,Slit:Jones1989,Slit:Affleck95, Slit:Borda07,Slit:lechtenbergAnders14,Slit:lechtenberg_2016_dimer}
to two impurities which were separated by a distance $R$.
The conduction band is divided into two conduction bands, one with even and one with odd parity symmetry, whose effective densities of states (DOSs) incorporated the spatial extension.
In the following, we briefly summarize this procedure for the two impurity Anderson model (TIAM).

The Hamiltonian of the TIAM can be separated into three parts $H=H_\mathrm{c} + H_\mathrm{D} + H_\mathrm{I}$.
$H_c$ contains the conduction band $H_c=\sum_{\vec{k},\sigma}\epsilon_{\vec{k}}c^{\dagger}_{\vec{k},\sigma}c^{\p}_{\vec{k},\sigma}$
where $c^\dagger_{\vec{k},\sigma}$ creates an electron with spin $\sigma$ and momentum $\vec{k}$.
$H_\mathrm{D}$ comprises the contribution of the impurities
\begin{align}
	H_\mathrm{D}=&\sum_{j,\sigma} E_j d^\dagger_{j,\sigma} d^{\phantom{\dagger}}_{j,\sigma} + U \sum_j n_{j,\uparrow}n_{j,\downarrow} 
		+ \frac{t}{2} \sum_\sigma \left( d^\dagger_{1,\sigma} d^{\phantom{\dagger}}_{2,\sigma} + d^\dagger_{2,\sigma} d^{\phantom{\dagger}}_{1,\sigma} \right), \label{eq:TIAM:H_d}
\end{align}
with $d^\dagger_{j,\sigma}$ creating an electron on impurity $j$ with the level energy $E_j$, $n_{j,\sigma}=d^\dagger_{j,\sigma} d^{\p}_{j,\sigma}$, 
the Coulomb repulsion $U$ and a tunneling $t$ between the impurities.
The interaction between the conduction band and the impurities is given by
\begin{align}
	H_\mathrm{I}=&  \frac{V}{\sqrt{N}}\sum_{j\in\{1,2\}k,\sigma} c^\dagger_{\vec{k},\sigma} e^{\mathrm{i}\vec{k}\vec{R}_j} d_{j,\sigma} + \mathrm{h.c.} ,
\end{align}
where $V$ is the hybridization strength between the conduction band and the two impurities which are located at $\vec{R}_{1/2}=\pm\frac{\vec{R}}{2}$.
For simplicity we will consider the case particle-hole and parity symmetric case $E_1=E_2=-U/2$.

For the NRG, it is useful \cite{Slit:JayaprakashKrischnamurtyWilkins1981,Slit:Jones87,Slit:Jones88,Slit:Jones1989,Slit:Affleck95, Slit:Borda07,Slit:lechtenbergAnders14,Slit:lechtenberg_2016_dimer} 
to include spatial dependency into the two orthogonal energy-dependent even ($e$) and odd ($o$) parity eigenstate field operators
\begin{align}
c_{\epsilon,\sigma,e/o } =& \sum_{\vec{k}} \delta(\epsilon-\epsilon_{\vec{k}}) c_{\vec{k}\sigma}  
			  \frac{\left( e^{+ i\vec{k}\vec{R}/2} \pm e^{- i\vec{k}\vec{R}/2} \right)}{N_{e/o}(\epsilon, \vec{R})\sqrt{N\rho_{c}(\epsilon)}}.
\end{align}
Here $\rho_c(\epsilon)$ is the DOS of the original conduction band and the dimensionless normalization functions are defined as
\begin{subequations}
\begin{align}
N^2_{e}(\epsilon,\vec{R})= &\frac{4}{N\rho_{c}(\epsilon)} \sum_{\vec{k}} \delta(\epsilon-\epsilon_{\vec{k}}) \cos^2\left(\frac{\vec{k}\vec{R}}{2}\right) \\
N^2_{o}(\epsilon,\vec{R})=& \frac{4}{N\rho_{c}(\epsilon)} \sum_{\vec{k}} \delta(\epsilon-\epsilon_{\vec{k}}) \sin^2\left(\frac{\vec{k} \vec{R}}{2}\right). \label{eq:NormFactor}
\end{align}
\end{subequations}
such that $c_{\epsilon,\sigma,e/o}$ fulfill the standard anticommutator relation $\{c^{\phantom{\dagger}}_{\epsilon,\sigma,p},c^\dagger_{\epsilon',\sigma',p'}\}=\delta_{\sigma,\sigma'}\delta_{p,p'}\delta(\epsilon-\epsilon')$.
If we also introduce even and odd parity combinations for the orbitals
\begin{align}
	d_{e/o,\sigma} =\frac{1}{\sqrt{2}}\left( d_{1,\sigma} \pm d_{2,\sigma} \right)
\end{align}
$H_\mathrm{D}$ and $H_\mathrm{I}$ are given by
\begin{align}
	H_\mathrm{D} =& \sum_{p=\{{e,o}\},\sigma} E_p d^\dagger_{p,\sigma} d^{\phantom{\dagger}}_{p,\sigma} + \frac{U}{2} \sum_p n_{p,\uparrow} n_{p,\downarrow} 
        + \frac{U}{4}\sum_{\sigma,\sigma'} n_{e,\sigma}n_{o,\sigma'} - U \vec{S}_e\vec{S}_o 
        + \frac{U}{2} \left( d^\dagger_{e,\uparrow} d^\dagger_{e,\downarrow} d^{\phantom{\dagger}}_{o,\downarrow} d^{\phantom{\dagger}}_{o,\uparrow} + \mathrm{h.c.} \right) \\
	H_\mathrm{I}=& \frac{V}{\sqrt{2}} \sum_{\sigma} \int d\epsilon \sqrt{\rho_c(\epsilon)}\left\{ N_{e}(\epsilon,\vec{R} ) c^\dagger_{e,\sigma} (\epsilon) d^\pd_{e,\sigma} \right. 
		        \left. + N_{o}(\epsilon,\vec{R} ) c^\dagger_{o,\sigma}(\epsilon) d^\pd_{o,\sigma} + \mathrm{h.c.}  \right\} \\
		    =& \sum_{\sigma} \left\{ \sqrt{\frac{\Gamma_{e}(\epsilon,{R})}{\pi}} c^\dagger_{e,\sigma} (\epsilon) d^\pd_{e,\sigma} \right. 
		      \left. + \sqrt{\frac{\Gamma_{o}(\epsilon,{R})}{\pi}}c^\dagger_{o,\sigma}(\epsilon) d^\pd_{o,\sigma} + \mathrm{h.c.} \right\}. \label{eq:H_I_HybFunc}
\end{align}
Here we defined the local spin operator $\vec{S}_{e/o}= \frac{1}{2} \sum_{\alpha,\beta} d^\dagger_{e/o,\alpha} \vec{\sigma}_{\alpha,\beta} d^{\pd}_{e/o,\alpha}$ with $\vec{\sigma}$ being Pauli matrices, 
and the even and odd energy level is given by $E_{e/o}=E\pm t/2$.
The hybridization functions in \eqref{eq:H_I_HybFunc} are defined as $\Gamma_{e/o}(\epsilon,{R})=  \pi V^2 \rho_c({\epsilon}) \frac{N^2_{e/o}(\epsilon,\vec{R})}{2}$
and inserting a 1D linear dispersion $\epsilon(k)=v_\mathrm{F}(|k|-k_\mathrm{F})$ yields \cite{Slit:Borda07,Slit:lechtenbergAnders14}
\begin{align}
	\Gamma_{e/o}^{\mathrm{1D}}(\epsilon,{R})=\Gamma_0\left( 1 \pm \cos\left[ k_\mathrm{F} R ( 1 + \frac{\epsilon}{D})\right] \right) \label{Seq:HybridizationFunctions_1D}
\end{align}
with $\Gamma_0=\pi\rho_0 V^2$, the half bandwidth $D$, the constant DOS of the original conduction band $\rho_0=1/2D$, $k_\mathrm{F}=\pi/2a$ and the lattice constant $a$.
The hybridization functions $\Gamma^\mathrm{1D}_{e/o}(\epsilon,{R})$ are depicted for a 1D linear dispersion in Fig.\ \ref{fig:DOS}
for two different distances $k_\mathrm{F}R=\pi,2\pi$.
The hybridization function $\Gamma^\mathrm{1D}_{e}(\epsilon,{R})$ of the even conduction band exhibits a gap for distances $k_\mathrm{F}R=(2n+1)\pi$ 
whereas the one of the odd band $\Gamma^\mathrm{1D}_{o}(\epsilon,{R})$ has a gap for $k_\mathrm{F}R=2n\pi$.

\section{Discontinuity in the correlation function $\langle \vec{S}_1 \vec{S}_2 \rangle$}

In order to understand the properties of the protected QCP and its differences to the previous known QCP \cite{Slit:Jones88,Slit:Jones1989,Slit:Affleck95,Slit:Galpin2005} 
it is useful to consider the effective low temperature Hamiltonian for the large $U$ limit.
Via a Schrieffer-Wolff transformation \cite{Slit:SchriefferWolff1966, Slit:ZitkoBonca2006} one obtains the two impurity Kondo model (TIKM) \cite{Slit:Jones87,Slit:Jones88,Slit:Jones1989,Slit:Affleck95}
\begin{subequations}
\begin{align}
	H_\mathrm{K,I}=& \frac{ J}{8} \int \int \; d\epsilon \; d\epsilon' \sqrt{\rho_c(\epsilon) \rho_c(\epsilon')} \sum_{\sigma\sigma'} \vec{\sigma}_{\sigma\sigma'} \nonumber \\
	   &\times\left[ (\vec{S}_1 \right.  \left.  +\vec{S}_2) \sum_p \left( N_p(\epsilon,R)N_p(\epsilon',R) c^\dagger_{\epsilon \sigma,p}c^{\pd}_{\epsilon' \sigma',p} \right) \right. 
	  \left.+ (\vec{S}_1  \right.  \left.  -\vec{S}_2) N_e(\epsilon,R)N_o(\epsilon',R) \left( c^\dagger_{\epsilon \sigma,e}c^{\pd}_{\epsilon' \sigma',o} 
	      + \mathrm{h.c.} \right) \right]  \label{eq:TIKM:H_I} \\
	  H_{K,D}  =& K \vec{S}_1\vec{S}_2. \label{eq:TIKM:H_d}
\end{align}
\label{eq:TIKM}
\end{subequations}
with $J=\frac{8V^2}{U}$ and $K=\frac{t^2}{U}$.

At low temperature, the last term of Eq.~\eqref{eq:TIKM:H_I}, which is proportional to $(\vec{S}_1 -\vec{S}_2)$, always vanishes since either $N_e(\epsilon,R)\to 0$ or $N_o(\epsilon,R)\to 0$ for $\epsilon \to 0$.
This term transfers ``parity'' from the impurity to the conduction band, therefore, $\langle \vec{S}_1\vec{S}_2\rangle$ may change continuously from a triplet with parity $+1$ to a singlet 
with parity $-1$ as long as this term is present.
However, since this term disappears at low energy scales, the correlation function has to change discontinuously at the QCP for a parity symmetric model.
This discontinuity is hence a consequence of parity conservation.

\section{Unstable intermediate fixed point}

\begin{figure}[t]
	\includegraphics[width=0.65\textwidth]{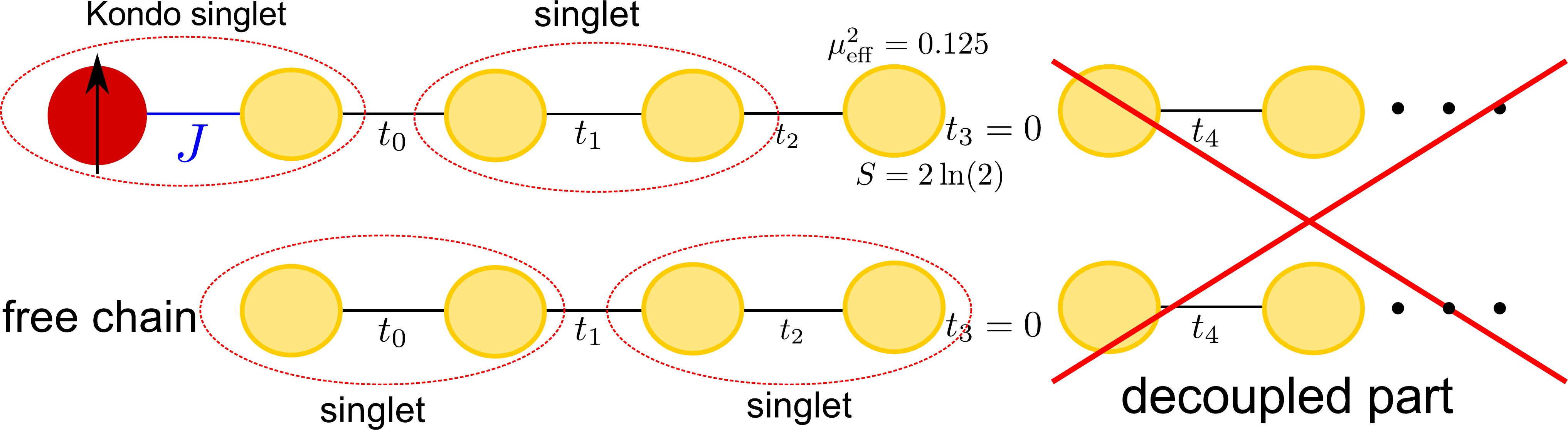}
	\caption{Schematic view of the gapped Wilson chain (top) with the impurity coupled to it and (bottom) without the impurity.
		The values $\mu^2_\mathrm{eff}=0.125$ for the magnetic moment and $S=2\ln(2)$ for the entropy compared to the free chain 
		are caused by the by a Wilson site which is not bound into a singlet.
	}
	\label{fig:Explanation_instable_fixpoint}
\end{figure}

To reveal that the unstable intermediate fixed point is just a feature of the Wilson chain for a pseudogap DOS
we recall that
impurity contributions are computed with the numerical renormalization group (NRG)
by calculating a quantity for the whole system, consisting of the impurity coupled to the bath, and subtracting the quantity of the system without impurity from it
\begin{align}
	A(T) = A_\mathrm{tot}(T) - A_\mathrm{free}(T) \label{eq:Impurity_contribution}
\end{align}
where $A_\mathrm{tot}(T)$ is the measured quantity of the whole system and $A_\mathrm{free}(T)$ the one without impurity.

The pseudogap in $\Gamma_{e/o}(\epsilon,{R})$ leads to a characteristic features of the tight-binding
matrixelements $t_n$ of the Wilson chain: at the energy scale $E_m\propto D\Lambda^{-m/2}$,
corresponding to the energy scale $E_{\rm gap}$ at which the pseudogap starts to develop, 
the matrixelement $t_m$ is strongly reduced, and for $n>m$, the matrixelements $t_n$
alternate in magnitude.
Since a fermionic DOS has negative and positive energy contributions, $m$ must be odd. The $t_n$ for even 
$n$ are large forming a binding (negative energy) and anti-binding orbital representing the  energy scale 
$E_n \propto D\Lambda^{-n/2}$ which are only weakly connected to orbitals of neigboring energy shells, i.\ e.\ $t_n$
is small for odd $n$.

In Fig.~\ref{fig:Explanation_instable_fixpoint}, a schematic view of the gapped Wilson chain (top) with the impurity coupled to it and (bottom) without impurity is shown. 
For clarity, we artificially set $t_m=0$ -- here depicted for $m=3$ -- and disconnect the rest of the chain from the problem. 
Furthermore, we consider the strong coupling limit $J\to\infty$ for a single band $s-1/2$ Kondo model.
This decoupled part is identical for the whole system and the free chain.  Since it  decouples from
the problem for $t_m=0$, these degrees of freedom do not contribute to Eq. \eqref{eq:Impurity_contribution}.

At each NRG iteration all eigenenergies are rescaled by a factor $\sqrt{\Lambda}$, with $\Lambda$ being the discretization parameter of the NRG.
Therefore, high energy states of the first part are renormalized to $\infty$ so that only the ground state of the first part of the Wilson chain for the iteration $N=m$ remains.

This ground state is a singlet, however,
the chain with impurity has one Wilson site that is not bound into a singlet, cf. Fig.~\ref{fig:Explanation_instable_fixpoint}. 
For a particle-hole symmetric pseudogap DOS, the last site contributes a free orbital, 
leading to an effective magnetic moment of $\mu^2_\mathrm{eff}=0.125$ and a entropy $S_{\rm imp}=2\ln(2)$ employing the definition \eqref{eq:Impurity_contribution}.
Therefore, the values for the magnetic moment and entropy at the unstable FP (Fig. 2 in the paper) are 
essentially contributions from the Wilson chain for a pseudogap DOS. This also illustrates, that the effective free moment
of the stable FP is generated mainly from the conduction electron degrees of freedom.

The ground state of the first part, however, is almost degenerated with other excited states depending on the effective renormalized Kondo coupling $J$.
The larger the renormalized coupling $J$ the smaller the energy difference between the ground state and those excited states.
This energy difference defines the energy scale at which the transition to the USK FP occurs so that the apparent degeneracy is lifted due to the rescaling of the energies and the system flows to the stable USK FP.

\section{Local moment of the impurity $\mu_\mathrm{loc}^2$}

\begin{figure}[t]
	\includegraphics[width=0.65\textwidth]{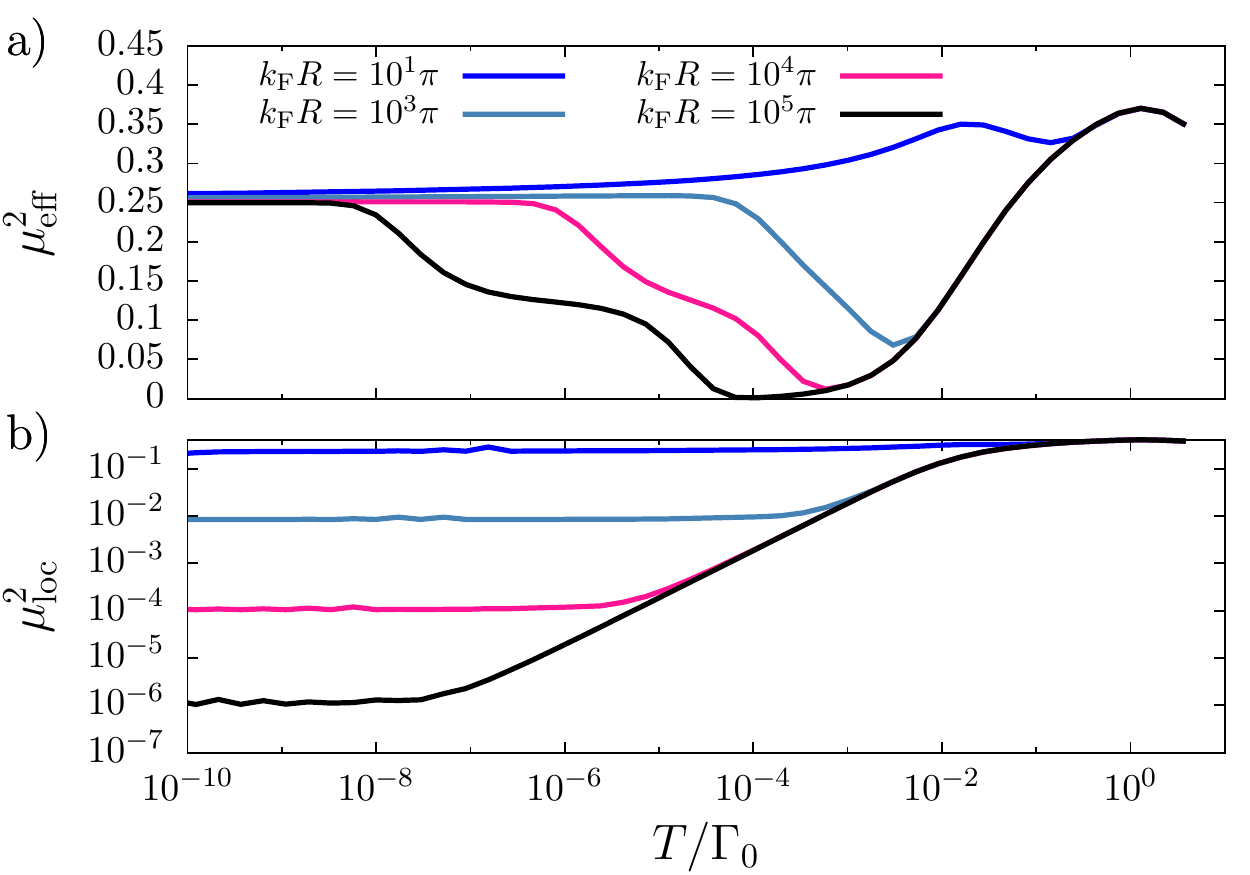}
	\caption{(a) Effective local magnetic moment $\mu_\mathrm{eff}^2$ and (b) the local moment of an impurity $\mu_\mathrm{loc}^2$, calculated by applying a small magnetic field $h_z=10^{-10}\Gamma_0$ to the impurity spin.
		In contrast to $\mu_\mathrm{eff}^2$, the impurity local moment $\mu_\mathrm{loc}^2$ shows no intermediate unstable fixed point.
	}
	\label{fig:Comparison_eff_loc}
\end{figure}

As discussed in the paper, the effective local magnetic moment $\mu_\mathrm{eff}^2(t)$, depicted in Fig.~\ref{fig:Comparison_eff_loc}a, 
reveals an intermediate unstable FP caused by the Wilson chain for a pseudogap DOS.
The value of $\mu^2_\mathrm{eff}=0.125$ is generated by the conduction band based on the mechanism discussed
in the previous section. 

In contrary to $\mu_\mathrm{eff}^2(T)$, the local response
$\mu^2_\mathrm{loc}(T) = T \lim_{h_z\to 0} \langle S^z_j\rangle/h_z$ 
demonstrate that the impurity spins remain screened \cite{Slit:Ingersent1998,Slit:Chowdhury2015} 
which can be seen in Fig.\ \ref{fig:Comparison_eff_loc}(b).

While $\mu^2_\mathrm{eff}(T)$ starts to increase
for decreasing temperatures corresponding to
the energy scale at which the gap occurs until it reaches the value $\mu^2_\mathrm{eff}=0.125$ in the regime of the unstable fixed point, the impurity spins 
are continued to be screened, and the 
local moment of the impurities $\mu^2_\mathrm{loc}(T)$
decrease linearly with decreasing $T$, 
since the local susceptibility $\chi_{\rm loc} = \lim_{h_z\to 0} \langle S^z_j\rangle/h_z$ has reached a constant value and shows the behavior of a Pauli-susceptibility characteristic for a Kondo screened impurity.
The screening of the impurity spins progresses until the underscreened Kondo fixed point (USK FP) point is reached at low temperatures,
because the impurity spins are only completely screened at  $T=0$ in the conventional Kondo problem.
There, the effective local magnetic moment takes the value $\mu^2_\mathrm{eff}=0.25$ and $\mu^2_\mathrm{loc}(T)$ reaches a very small but finite value
corresponding to a Curie-Weiss behavior of a free but strongly reduced magnetic moment.
$\mu^2_\mathrm{loc}(T)$ is a monotonically decreasing function with decreasing temperature
until the USK FP is reached as shown in Fig.\ \ref{fig:Comparison_eff_loc}(b).
Since $\mu^2_\mathrm{eff}(T)/\mu^2_\mathrm{loc}(T)=\mathrm{const}$ at the stable FP, there is very small but finite
overlap between the local magnetic moment and the free effective spin at the FP.

\section{Phase diagrams for $\mu_\mathrm{eff}^2$ and $S_\mathrm{imp}$}
\begin{figure}
	\includegraphics[width=0.65\textwidth]{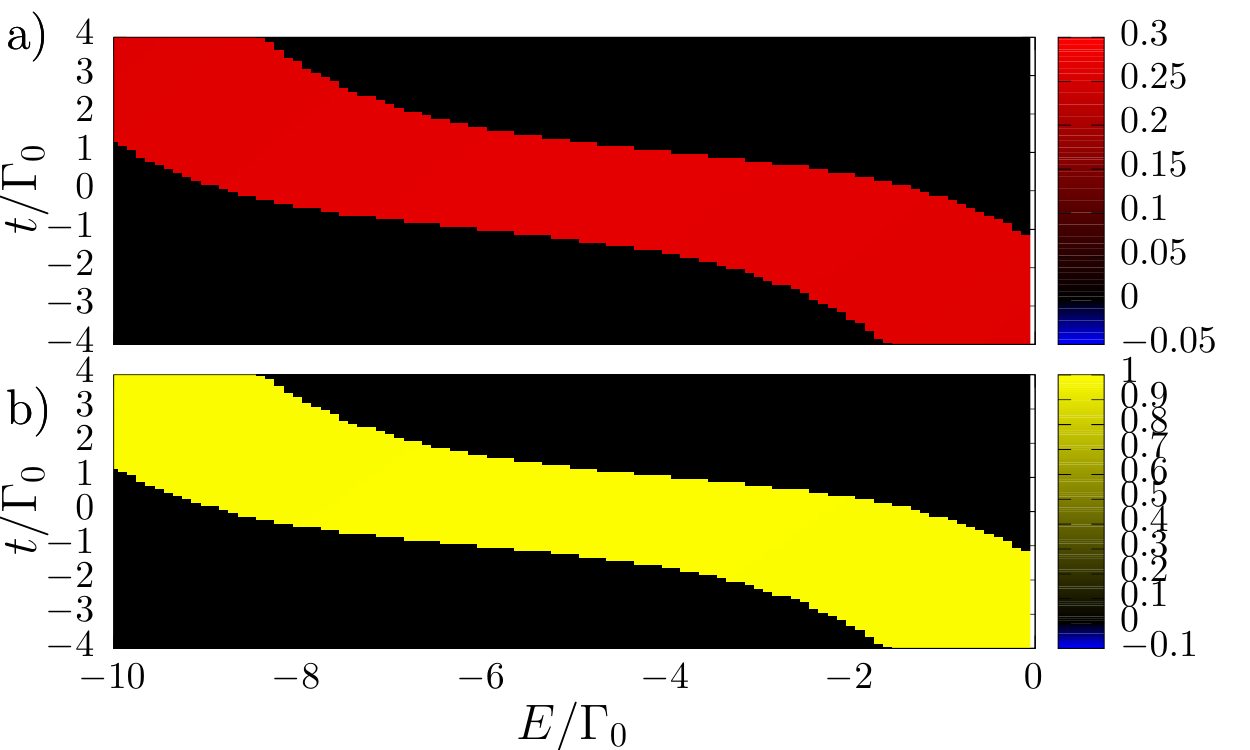}
	\caption{(a) The phase diagram of the effective local magnetic moment $\mu_\mathrm{eff}^2$ and (b) entropy $S_\mathrm{imp}/\ln(2)$ as a function of the tunneling $t$ and level energy $E$ 
	for $k_\mathrm{F}R=\pi$ and $U/\Gamma_0=10$.
	}
	\label{fig:phase_diagrams}
\end{figure}
Figure \ref{fig:phase_diagrams} shows the phase diagrams of the (a) effective local magnetic moment $\mu_\mathrm{eff}^2$ and (b) entropy $S_\mathrm{imp}/\ln(2)$ plotted against the tunneling $t$ and level energy $E$.
Inside a tube the system is in the doublet phase which can be seen on the fixed values of the local moment $\mu_\mathrm{eff}^2=0.25$ and entropy $S_\mathrm{imp}/\ln(2)=1$.
By changing the tunneling $t$ or level energy $E$ a transition to the singlet phase with $\mu_\mathrm{eff}^2=0$ and $S_\mathrm{imp}/\ln(2)=0$ can be evoked.
Unlike the correlation function $\langle\vec{S}_1\vec{S}_2\rangle$, which continuously vanish for large $|E|$ as shown in Fig. 4 in the main paper, 
the local magnetic moment and entropy can only take the two above mentioned values.
Note, however, that although it is still possible to evoke a transition from the singlet to the doublet phase for very large $|E|$ using an appropriate tunneling $t$, 
a Kondo effect would not occur anymore in the doublet phase since the coupling orbital would be empty or doubly occupied.

\section{Deviations from $R_n$}
\begin{figure}[t]
	\includegraphics[width=0.65\textwidth]{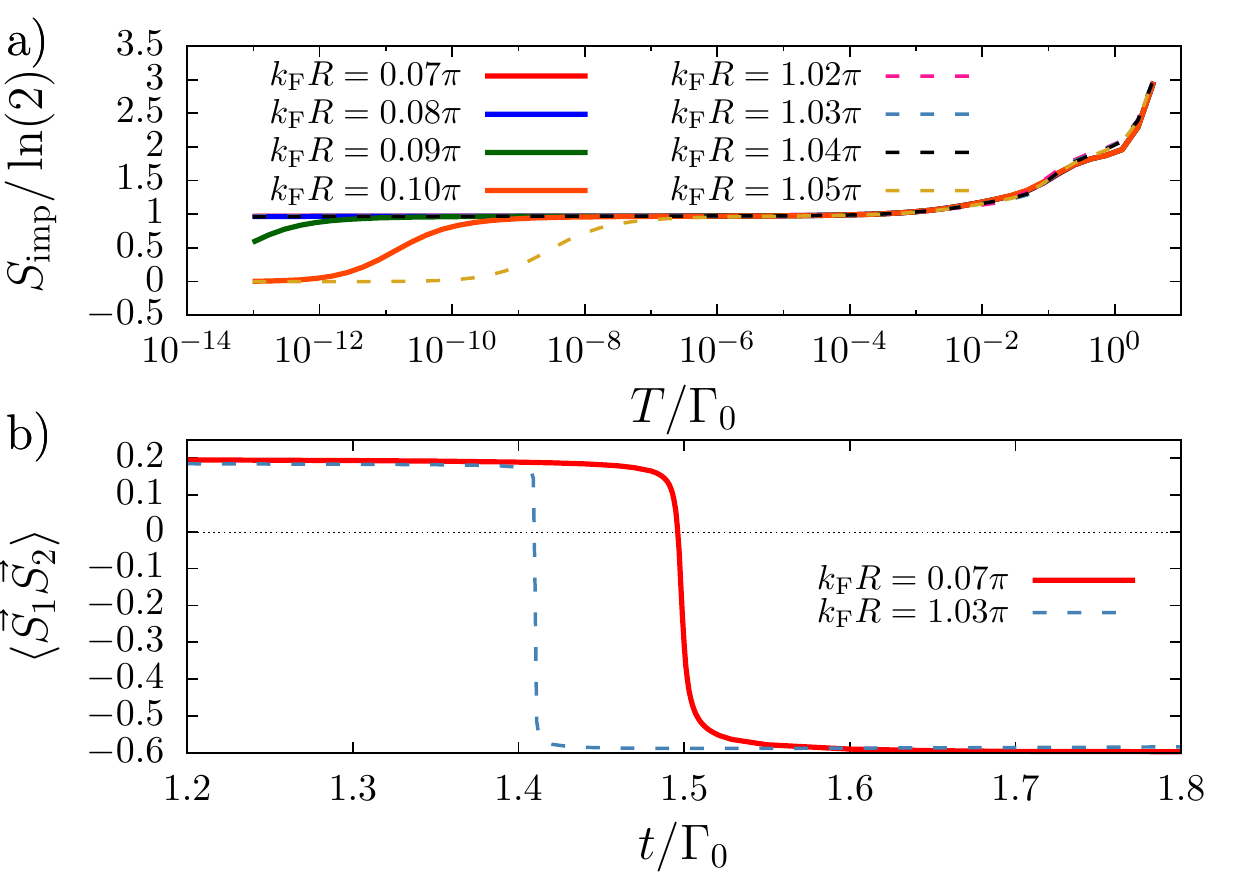}
	\caption{(a) The temperature dependent entropy for $t=0$ and distances that slightly deviate from $R_n=0$ (solid lines) and $R_n=1$ (dashed lines).
	(b) $\langle \vec{S}_1 \vec{S}_2 \rangle$ plotted against $t$ for the two distances $k_\mathrm{F}R=0.07 \pi$ and $k_\mathrm{F}R=1.03 \pi$ at finite temperature $T/\Gamma_0\approx10^{-7}$.
	}
	\label{fig:deviation_from_R}
\end{figure}
In Fig. \ref{fig:deviation_from_R}a the entropy is plotted against the temperature for distances that slightly deviate from $R_n$.
Although a small departure from the specific distances $R_n$ theoretically always leads to a singlet ground state and consequently to a vanishing entropy $S_\mathrm{imp}=0$ for $T \to 0$,
the system stays in the now unstable doublet fixed point for a wide temperature range if the departure is not to large.
For $k_\mathrm{F}R_n=0$ departures up to $\Delta R=0.10\pi/k_\mathrm{F}=0.20a$ and for $k_\mathrm{F}R_n=\pi$ departures up to $\Delta R=0.05\pi/k_\mathrm{F}=0.10a$, with $a$ being the lattice constant,
are still sufficient to detect a sharp change of $\langle \vec{S}_1 \vec{S}_2 \rangle$, $S$ and $\mu^2_\mathrm{eff}$ from the doublet to the singlet phase.
This can be seen in Fig. \ref{fig:deviation_from_R}b where $\langle \vec{S}_1 \vec{S}_2 \rangle$ is plotted against the tunneling $t$ 
for the two distances $k_\mathrm{F}R=0.07 \pi$ and $k_\mathrm{F}R=1.03 \pi$ at finite temperature $T/\Gamma_0\approx10^{-7}$.
Although the change of $\langle \vec{S}_1 \vec{S}_2 \rangle$ from the doublet to the singlet phase is now continuous at finite temperatures,
it happens on a very narrow energy scale $\Delta t \approx 0.05 \Gamma_0$.
As a result, experiments at finite temperatures will detect features of the QCP 
such as sharp changes in the magnetic properties of the system even for small departures from the specific distances $R_n$.

%

\end{document}